\documentstyle[12pt,epsf,epsfig]{article}
\def\beq{\begin{equation}}
\def\eeq{\end{equation}}
\def\bea{\begin{eqnarray}}
\def\eea{\end{eqnarray}}
\def\ba{\begin{array}}
\def\ea{\end{array}}

\def\tg{\tilde g}
\def\ty{\tilde y}
\def\th{\tilde h}

\def\calo{{\cal O}}

\newcommand{\NPB}[3]{\emph{ Nucl.~Phys.} \textbf{B#1} (19#2) #3}   
\newcommand{\PLB}[3]{\emph{ Phys.~Lett.} \textbf{B#1} (19#2) #3}   
\newcommand{\PRD}[3]{\emph{ Phys.~Rev.} \textbf{D#1} (19#2) #3}

\newcommand{\PR}[3]{\emph{ Phys.~Rep.} \textbf{#1} (19#2) #3}

\def\gappeq{\mathrel{\rlap {\raise.5ex\hbox{$>$}}
{\lower.5ex\hbox{$\sim$}}}}
\def\permil{$\%\raise.20ex\hbox{$_0$}}
\def\lappeq{\mathrel{\rlap{\raise.5ex\hbox{$<$}}
{\lower.5ex\hbox{$\sim$}}}}
\begin{document}
\topmargin -1.0cm
\oddsidemargin -0.8cm
\evensidemargin -0.8cm
\pagestyle{empty}
\begin{flushright}
UG-FT-111/99
\end{flushright}
\vspace*{5mm}
\begin{center}
{\Large\bf Unification through extra dimensions
}\\
\vspace{0.5cm}
{\Large\bf at two loops}\\
\vspace{2cm}
{\large\bf M. Masip}\\
\vspace{.8cm}
{{\it Dpt. F\'\i sica Te\'orica y del Cosmos}\\
{\it Universidad de Granada}\\
{\it E-18071 Granada, Spain}\\}
\end{center}
\vspace{2cm}
\begin{abstract}
The presence of an extra dimension of size $R\equiv M_c^{-1}$ 
introduces corrections of order $(\mu/M_c)\alpha$
to the gauge and Yukawa
couplings and {\it accelerates} their running at scales 
$\mu$ larger than $M_c$. This could result in a grand
unification scale $M_X\approx 20 M_c$. We study the corrections
at the two-loop level. We find corrections 
of order $(\mu/M_c)\alpha^2$ for the gauge couplings and 
of order $(\mu/M_c)^2\alpha^2$ for the Yukawa couplings.
Therefore, in the Yukawa sector one and 
two-loop contributions can be of the same order below $M_X$.
We show that in the usual scenarios the dominant gauge 
and Yukawa couplings are decreasing functions 
of the scale, in such a way that $(\mu/M_c)\alpha$
becomes approximately constant and two-loop contributions
introduce just a $30\%$ correction 
which does not increase with the scale.

\end{abstract}

\vfill

\eject
\pagestyle{empty}
\setcounter{page}{1}
\setcounter{footnote}{0}
\pagestyle{plain}


\section{Introduction}

String theory, the only known candidate for a 
fundamental theory including gravity,
is formulated in more than three spatial dimensions. 
Although the
extrapolation of the observed physics 
to higher energies suggests a very small size of 
the extra dimensions, of 
$\calo (M^{-1}_{Planck})$, it has been recently proposed 
an alternative framework where 
some of these dimensions could have a large size \cite{ark98}. 
In particular, a compact dimension of 
inverse radius $R^{-1}\equiv M_c=\calo(1\; {\rm TeV})$ 
where only the gauge bosons and the Higgs fields 
(but not quarks and leptons) propagate could define 
a viable scenario within the context of superstring theory
\cite{kak98a}.

The presence of such a compact dimension at low scales has
consequences on the unification of the gauge couplings 
of the standard model \cite{die98}. In the minimal supersymmetric 
extension of the standard model (MSSM) the observed values 
are a consequence of the different running 
from the grand unification scale. Obviously,
this picture would be modified by a fifth dimension in the
TeV region. First, we note that 5-dimensional (5D)
field theories are not renormalizable 
and must be taken as effective
theories valid only below a more fundamental mass scale, 
the string (unification) scale $M_X$. 
Second, in order to describe the effective theory we observe that 
the momentum along a compact dimension is quantized 
in units of $M_c$. This will translate into an infinite tower
of Kaluza-Klein (KK) modes associated to any field that penetrates
the extra dimension. Below $M_X$ it is possible to 
approximate the theory by a 4D truncated theory 
where the high energy modes have been cutoff. 
In particular, the 
authors in \cite{die98} show that the 
one-loop corrections to the gauge couplings in the 5D theory
almost coincide with the running in the truncated theory.
The regularization
dependence, which would describe the KK threshold effects, has
been discussed in \cite{kub99}.

The KK approach provides a simple framework to understand the
running of the couplings. A degree of freedom will be 
included in the loop only at scales larger 
than its mass. Since at a given scale
$\mu > M_c$ the number of KK modes with masses lighter than $\mu$
is $N(\mu)=2\mu/M_c$, the one-loop renormalization group
equations (RGEs) become
\begin{equation}
{{\rm d} \alpha^{-1}_i\over {\rm d}\ln\mu}  = -{1\over 2\pi} b_i 
\rightarrow
{{\rm d} \alpha^{-1}_i\over {\rm d}\ln\mu}  = -{1\over 2\pi} b_i 
{2\mu\over M_c} \;, 
\end{equation}
where $\alpha_i=g^2_i/(4\pi)$.
The first equation gives a linear dependence of $\alpha^{-1}_i$ with
$\ln \mu$, whereas the second one predicts a much faster exponential
(or linear versus $\mu$) behaviour.

In addition, the $b_i$ coefficients will change as well
due to the different field content of a superfield in five
dimensions. Actually, the 5D content can be accommodated
in complete 4D hypermultiplets of $N=2$ 
supersymmetry (SUSY) \cite{soh85}. 
Each gauge boson will come with its gaugino plus a 
fermion and a scalar in the adjoint representation of 
the gauge group, defining a complete vector hypermultiplet
(equivalent to a vector plus a chiral supermultiplets).
For the matter fields, a chiral hypermultiplet will 
include a vectorlike pair of chiral superfields. In 
this way, the two Higgs doublets of the MSSM could in
principle be different components of the same 5D field.

Since an extra dimension at a low scale $M_c$ would
increase exponentially the evolution rate of the 
couplings, gauge unification could occur at 
much lower energies than in standard scenarios.  
Unification would not necessarily introduce 
a hierarchy $M_X/M_{EW} = \calo (10^{14})$: if
$M_c\approx M_{EW}$, then $M_X/M_{EW} = \calo (14)$.
The possibility of unification at a scale $M_X$ 
very close to $M_c$ has been explored recently by 
different authors \cite{die98,ghi98}. 
Their one-loop analysis suggests 
gauge and also Yukawa coupling unification in 
several definite models.
The behaviour that they obtain 
above $M_c$ is similar to the standard
one with the changes $\ln\mu\rightarrow \mu$ and
$b_i\rightarrow \tilde b_i$. However, there are 
important differences in the running when one 
goes beyond the one-loop level which cannot be
accounted with just these substitutions. 
In particular, whereas the 4D RGEs include
at one loop the corrections of type $(\alpha\ln\mu)^2$,
here the calculation will not include the corrections 
$(\alpha \mu)^2$. Here the parameter
in the loop expansion is $\alpha \mu/M_C$, not
just $\alpha$. Although the fields couple with 
strength $\alpha$, their number increases linearly
with the energy and the analysis remains perturbative
only if the product $\alpha \mu/M_c$ is small. 
Then, even
if $M_X\approx 20 M_c$ higher order corrections may
be already of order one.

In this paper we analyze the corrections to the
one-loop $\beta$-functions of the gauge and the Yukawa
couplings in models with an extra dimension\footnote{
These corrections to the gauge couplings are 
in addition to the ones from higher-dimensional 
operators \cite{che99} or massive string modes.}.
We show that whereas
the gauge sector is protected by the symmetries, in
the Yukawa sector the one-loop result is 
insufficient, since it is modified 
by higher order corrections proportional to 
$(h_t^2 \mu/M_c)^n$ and $(g_3^2 \mu/M_c)^n$. 
Due to these corrections $h_t$ could evolve fast
towards large values, which would affect the (two-loop)
evolution rate and the perturbative unification
of the gauge couplings. In the models
analyzed here, however, we observe the opposite 
behaviour due to the negative sign of 
the two-loop $\beta$-function,
and $h_t$ introduces little change in the
pattern of gauge unification obtained at one loop.

\section{Two-loop running}

Let us start defining the 5-dim extension of the MSSM.
The gauge and Higgs fields will live in the 5D bulk,
whereas the fermions are localized on 4D boundaries and
do not have KK excitations. At $M_c$ we  
find a chiral superfield $\Sigma^A$ for each gauge
superfield $V^A$, whereas in the chiral sector there are
pairs ($H,H'$) of Higgs 
superfields. Each of these fields has an infinite tower of 
KK modes of mass $nM_c$.
The compactification of the fifth dimension 
must be such that the zero modes define the MSSM, with no
massless $\Sigma^A$ fields. This usually forces the presence
of more than one pair of 5D Higgs fields. To simplify we will
consider only one pair of Higgs doublets \cite{die98} and
comment later on a case with two pairs \cite{pom98}. 
We will also assume that {\it (i)}
the couplings of the 4D chiral fields to the KK 
excitations coincide with their couplings to the zero modes
and {\it (ii)} the number of excitations is the usual in a circle.
This is not the case for the compactification on the 
segment $S^1/Z_2$ \cite{pom98}, with half the resonances in the
circle and with the couplings to the heavy modes increased  
by a factor of $\sqrt {2}$. Our conclusions, however, 
will be the same in this scenario.

The tower of chiral superfields $\Sigma^A$ couples to 
the chiral Higgses with an effective superpotential of type
$W \supset \sqrt{2} g {T^{Ai}}_j\; \Sigma^A H_i H'_j$.
All the interactions of the quark and lepton fields break
the underlying $N=2$ SUSY (their couplings with
the heavy Higgs and vector modes do not conserve the momentum 
along the fifth dimension).

In a general $N=1$ SUSY model with superpotential
$W= {1\over 6}  y_{ijk} \Phi_i\Phi_j\Phi_k$
and gauge symmetry $G$ the $\beta$-functions
($\beta_x\equiv {{\rm d} x\over {\rm d}\ln\mu}$,
with $x$ a generic gauge or Yukawa coupling) can be 
deduced from the vector and chiral selfenergies. At one
loop they read \cite{jon84}
\begin{eqnarray}
\beta^{(1)}_{\tg} &=& \tg^3 Q\; ;
\nonumber\\
\beta^{(1)}_{\ty_{ijk}} &=& \ty_{ijp}{\gamma^{(1)p}}_k+ 
(k\leftrightarrow i) + (k\leftrightarrow j)\, ,
\label{beta}
\end{eqnarray}
where $\tg=g/(4\pi)$, $\ty_{ijk}=y_{ijk}/(4\pi)$ and
\begin{eqnarray}
Q &=& T(\Phi) - 3 C(V)\;;
\nonumber\\
{\gamma^{(1)i}}_j &=& P^i_{\; j} = {1\over 2}
\ty^{ikl} \ty_{jkl}-2 \tg^2 C(\Phi)\delta^i_{\; j}\, .
\label{beta0}
\end{eqnarray}
The group constants are $T(\Phi)\delta_{AB}= {\rm Tr}(T^AT^B)$,
$C(V)\delta_{AB} = f_{ACD}f_{BCD}$ and 
$C(\Phi)\delta^i_{\; j}=(T^A T^A)^i_{\; j}$.
In the model under study 
we have (see Fig.~1)
$Q=T(H)+T(H')+T(\Sigma)-3C(V)=2T(H)-2C(V)$, 
$P^H_{\; H}=P^{H'}_{\; H'}=0$ (due to the cancellation of 
$V$ and $\Sigma$ contributions), and the MSSM value
for the third fermion 
generation $P^f_{\; f}$ ($f=Q,t^c,b^c,L,\tau^c$).
Including the sum over all KK modes of mass below $\mu$ 
we obtain 
\begin{eqnarray}
\beta^{(1)}_{\tg_1} &=& {3\over 5}\;\tg^3_1\; 
{2\mu\over M_C}\; ;
\nonumber\\
\beta^{(1)}_{\tg_2} &=& -3\;\tg^3_2\; 
{2\mu\over M_C}\; ;
\nonumber\\
\beta^{(1)}_{\tg_3} &=&-6\; \tg^3_3\; 
{2\mu\over M_C}
\label{beta11}
\end{eqnarray}
and 
\begin{eqnarray}
\beta^{(1)}_{\th_t} &=& \th_t\; 
\Big(3\th^2_t+\th^2_b-{17\over 30}\tg_1^2-{3\over 2}\tg_2^2
-{16\over 3}\tg_3^2\Big)\;{2\mu\over M_C}\; ;
\nonumber\\
\beta^{(1)}_{\th_b} &=& \th_b\; 
\Big(3\th^2_b+\th^2_t-{1\over 6}\tg_1^2-{3\over 2}\tg_2^2
-{16\over 3}\tg_3^2\Big)\;{2\mu\over M_C}\; ;
\nonumber\\
\beta^{(1)}_{\th_\tau} &=& \th_\tau\; 
\Big(3\th^2_\tau-{3\over 2}\tg_1^2-{3\over 2}
\tg_2^2\Big) \;{2\mu\over M_C}\; ,
\label{beta12}
\end{eqnarray}
where $\th_t\equiv \ty_{Qt^cH'}$, $\th_b\equiv \ty_{Qb^cH}$
and $\th_\tau\equiv \ty_{L\tau^cH}$.

The generic two-loop $\beta$-functions can be written 
\cite{jon84}
\begin{eqnarray}
\beta^{(2)}_{\tg} &=& 2\tg^5C(V)Q-2g^3r^{-1}C(\Phi)
\delta^i_{\; j} P^j_{\; i}\; ;
\nonumber\\
{\gamma^{(2)i}}_j &=& -\Big[ \ty^{imp}\ty_{jmn}+
 2g^2C(\Phi)\delta^p_{\; j}\delta^i_{\; n} \Big] P^n_{\; p}
+2g^4C(\Phi)\delta^i_{\; j}Q\, ,
\label{beta20}
\end{eqnarray}
where $r=\delta_{AA}$ is the dimension of the group. 

In our framework the dominant two-loop correction
to the gauge couplings 
would be proportional to $(2\mu/M_C)^2$, coming
from two-point diagrams with exchange of heavy KK 
modes only. However, we have 
$P^H_{\; H}=P^{H'}_{\; H'}=0$ and 
$P^{\Sigma^A}_{\; \Sigma^B}=2g^2[T(H)-
C(\Sigma) ] \delta_{AB} =g^2 Q \delta_{AB}$,
and the leading contribution to $\beta^{(2)}_{\tg}$ vanishes
\cite{kak99}. 
This cancellation reflects the fact that a model with
$N=2$ SUSY has to be renormalized only at the one-loop level,
{\it i.e.}, the one-loop counterterms render the theory 
finite at any order of perturbation \cite{jon84,how83}.

The subleading two-loop correction, proportional to
$2\mu/M_C$, comes from diagrams with intermediate
chiral fermions or zero modes of gauge or Higgs fields.
This correction does not vanish because the 
underlying $N=2$ SUSY is broken by the interactions of 
the fields in the loops.
It can be written
\begin{eqnarray}
\beta^{(2)}_{\tg} &=& 2\tg^5C(V) \Big[ 3 \Big( T(H)+T(H')\Big) 
+T(f) + C(\Sigma) - 9 C(V) \Big] 
\nonumber\\
&& -2g^3r^{-1} C(H) \Big[ (4-6) g^2 C(H) d(H) + {1\over 2} y^{Hff'}
y_{Hff'} + (H\leftrightarrow H') \Big]
\nonumber\\
&& -2g^3 C(\Sigma) 2 g^2 \Big[ 2T(H) - C(\Sigma) \Big] 
\nonumber\\
&& -2g^3r^{-1} C(f) \Big[ {1\over 2} y^{fkl}
y_{fkl} - 2 g^2 C(f) d(f) \Big] \, ,
\label{beta21}
\end{eqnarray}
where $d(\Phi)$ is the dimension of the $\Phi$ representation
and $f$ stands for the chiral fermions.
The terms in the first line in Eq.~(\ref{beta21}) come from
the diagrams in Fig.~2a \cite{par84}. 
The diagrams with exchange of 
$(H,V)$, $(H',V)$
and $V$ fields have a factor of three, since there are three 
different
ways to distribute the momentum along the fifth dimension 
({\it i.e.}, the KK modes) among
the fields in the internal loops. For the diagrams with 
$(f,V)$ and $(\Sigma,V)$ in the loops there is only one 
possible configuration, corresponding to the exchange of a 
tower of $V$ and $\Sigma$ fields, respectively.
The contributions in the second, third and fourth
line in Eq.~(\ref{beta21}) are due to the diagrams in 
Figs.~2b,2c,2d, respectively. 
The diagrams involving $(H,\Sigma)$
have a factor of two and the ones with intermediate $(V,H)$ fields
have a factor of 3. The remaining diagrams have only one possible 
configuration of KK modes. We obtain
\begin{eqnarray}
\beta^{(2)}_{\tg_1} &=& \tg_1^3 \Big( {199\over 25} \tg^2_1 +
{27\over 5} \tg^2_2 + {88\over 5} \tg^2_3 - {26\over 5} \th^2_t
- {14\over 5} \th^2_b- {18\over 5} \th^2_\tau \Big) {2\mu\over M_C}
\nonumber\\
\beta^{(2)}_{\tg_2} &=& \tg_2^3 \Big( {3\over 2} \tg^2_1 +
\tg^2_2 + 24 \tg^2_3 - 6 \th^2_t
- 6 \th^2_b - 2 \th^2_\tau \Big) {2\mu\over M_C}
\nonumber\\
\beta^{(2)}_{\tg_3} &=& \tg_3^3 \Big( {11\over 5} \tg^2_1 +
9 \tg^2_2 -40 \tg^2_3 - 4 \th^2_t
- 4 \th^2_b \Big) {2\mu\over M_C}\, .
\label{betag22}
\end{eqnarray}

In the Yukawa sector the situation is completely different.
All the Yukawa interactions
break the $N=2$ SUSY structure present in the bulk
and allow power-law corrections proportional to $(2\mu/M_C)^2$.
These can be especially important if they involve top quark 
interactions, since $\th_t^2 (2\mu/M_C)$ may be of $\calo (1)$ 
below the unification scale $M_X$.

We obtain that the leading corrections
cancel in the Higgs selfenergies: ${\gamma^{(2)H}}_H=
{\gamma^{(2)H'}}_{H'}=0$. For a chiral fermion $f$
the generic expression in (\ref{beta20}) becomes 
\begin{equation}
{\gamma^{(2)f}}_f = - \th_{f}^2 P^{f^c}_{\; f^c}
-2g^2C(f) P^f_{\; f}
+2g^4C(f)\Big[ 2T(H)+C(\Sigma)-3C(V)\Big] \, ,
\label{cse2}
\end{equation}
where the three contributions have their origin in the 
diagrams in Figs.~3a,3b,3c, respectively \cite{wes84}. 
In particular, 
for the top quark we find
\begin{eqnarray}
{\gamma^{(2)t}}_t &=& \Big( {31\over 900}\tg_1^4-{9\over 4}\tg_2^4-
{80\over 9}\tg_3^4+{1\over 10}\tg_1^2\tg_2^2+{8\over 45}\tg_1^2
\tg_3^2+{8}\tg_2^2\tg_3^2+
\nonumber\\
&&{1\over 2}\tg_1^2\th_t^2-{3\over 2}
\tg_2^2\th_t^2+{1\over 10}\tg_1^2\th_b^2-{3\over 2}\tg_2^2\th_b^2
-2\th_t^4-2\th_b^4\Big)
\Big({2\mu\over M_C}\Big)^2\; ;
\nonumber\\
{\gamma^{(2)t^c}}_{t^c} &=& \Big({184\over 225}\tg_1^4-
{80\over 9}\tg_3^4+{128\over 45}\tg_1^2\tg_3^2-
\tg_1^2\th_t^2+{3}\tg_2^2\th_t^2-2\th_t^4-2\th_t^2\th_b^2\Big)
\Big({2\mu\over M_C}\Big)^2\; .
\label{betatop}
\end{eqnarray}
Analogous expressions can be found for the bottom quark and 
the tau lepton. 
The dominant terms in the two-loop Yukawa $\beta$-functions are
then
\begin{eqnarray}
\beta^{(2)}_{\th_t} &=& \th_t\Big( {767\over 900}\tg_1^4
-{9\over 4}\tg_2^4-
{160\over 9}\tg_3^4+{1\over 10}\tg_1^2\tg_2^2+{136\over 45}\tg_1^2
\tg_3^2+{8}\tg_2^2\tg_3^2-
\nonumber\\
&&{1\over 2}\tg_1^2\th_t^2+{3\over 2}
\tg_2^2\th_t^2+{1\over 10}\tg_1^2\th_b^2-{3\over 2}\tg_2^2\th_b^2
-4\th_t^4-2\th_b^4-2\th_t^2\th_b^2\Big)
\Big({2\mu\over M_C}\Big)^2\; ;
\nonumber\\
\beta^{(2)}_{\th_b} &=& \th_b\Big( {167\over 900}\tg_1^4
-{9\over 4}\tg_2^4-
{160\over 9}\tg_3^4+{1\over 10}\tg_1^2\tg_2^2+{8\over 9}\tg_1^2
\tg_3^2+{8}\tg_2^2\tg_3^2+
\nonumber\\
&&{1\over 2}\tg_1^2\th_t^2-{3\over 2}
\tg_2^2\th_t^2-{1\over 10}\tg_1^2\th_b^2+{3\over 2}\tg_2^2\th_b^2
-2\th_t^4-4\th_b^4-2\th_t^2\th_b^2\Big)
\Big({2\mu\over M_C}\Big)^2\; ;
\nonumber\\
\beta^{(2)}_{\th_\tau} &=& \th_\tau\Big( {303\over 100}\tg_1^4
-{9\over 4}\tg_2^4+{9\over 10}\tg_1^2\tg_2^2+
{9\over 10}\tg_1^2\th_\tau^2+{3\over 2}\tg_2^2\th_\tau^2
-4\th_\tau^4\Big)
\Big({2\mu\over M_C}\Big)^2\; .
\label{betah2}
\end{eqnarray}

\section{Numerical results}

For the gauge couplings, the two loops introduce corrections 
$\calo(\tg^2_i+\th^2_i)$ to the one-loop result (the tilde 
stands for the coupling over $4\pi$). 
For example, keeping only the top and the
color couplings we have $\beta^{(2)}_{\tg_3} /
\beta^{(1)}_{\tg_3} \approx 6.6 \tg^2_3 + 0.6 \th^2_t$.
In the Yukawa sector the two-loop corrections are
$\calo[(\tg^2_i+\th^2_i)(2\mu/M_c)]$. In particular, for
$h_t$ we have
$\beta^{(2)}_{\th_t} / \beta^{(1)}_{\th_t} \approx (2\mu/M_c) 
(-17.7 \tg_3^4-4\th_t^4)/(-5.3 \tg_3^2+3\th_t^2)$.
For an initial value $\th_t = \tg_3 = 1/(4\pi)$,
$\beta^{(2)}_{\th_t} / \beta^{(1)}_{\th_t} = 1$ for
$\mu\approx 9 M_c$.

This estimate shows that the unification of the gauge and 
Yukawa couplings at $\approx 20 M_c$ cannot be stablished 
from just a one-loop calculation, it requires the analysis
of higher order corrections. It could happen, for example, 
that the two-loop corrections 
increase the value of $h_t$ with the scale 
and make it very large at energies 
just above $9M_c$. This would also affect the evolution rate 
of the
gauge couplings and could spoil de gauge unification obtained
at one loop.

In the particular scenario under study, however, we 
obtain the opposite effect. At $M_c$ the one-loop 
$\beta$-functions for 
$\tg_3$ and $\th_t$ are both negative, which retards the 
scale where the one and two-loop $\beta$-functions 
become 
of the same order. In addition, the two-loop contribution
to $\beta_{\th_t}$ is also negative, further decreasing $\th_t$
and pushing up the range of energies where 
$\beta^{(2)}_{\th_t} / \beta^{(1)}_{\th_t}$ is small.
We plot in Fig.~4 the numerical running 
(including all the couplings)  
for $\tan \beta = 2$ and $M_c=1$ TeV. Dashes correspond
to the one-loop running, so one can see graphically the effect
of second order corrections. At $\mu\approx 10$ TeV two-loop 
effects account for a $25\%$ of the $\beta$-functions (see Fig.~5). 
This relative value of $\beta^{(2)}_{\th_2}$, however, does not 
increase since the linear behaviour of $\beta^{(2)}_{\th_2}$
with $\mu$ is dumped by the
running of $\th_t$ to smaller values 
dictated by the sign of its $\beta$-function.

We have also deduced the RGE in the model compactified in
the orbifold $S^1/Z_2$. This model includes just one mode
at each KK level, two pairs of Higgs $(H,H')$ KK towers, and a 
factor of $\sqrt{2}$ in the couplings of the chiral fermions
to the massive KK modes. Gauge unification is lost but the
behaviour of two-loop corrections is completely analogous.
We obtain (see Fig.~6) also negative values for 
$\beta^{(1)}_{\th_t}$ and $ \beta^{(2)}_{\th_t}$, which 
again avoids {\it large} couplings at scales just above
$\mu\approx 10 M_c$.

\section{Conclusions} 

We have studied whether in models with a compact dimension
the low-energy value of the 
gauge and Yukawa couplings can be related perturbatively
with the value at a scale $\mu > M_c$. In particular,
we have calculated the two-loop corrections in a model
with gauge unification at $M_X\approx 20 M_c$ and also
in the model compatified on $S^1/Z_2$.

For the gauge couplings
we obtain that the underlying $N=2$ SUSY in the
KK sector cancels the leading contributions in
the two-loop $\beta$-functions. In consequence,
$\beta^{(2)}_{g_i} / \beta^{(1)}_{g_i}=
\calo(\alpha/4\pi)$, where $\alpha$ refers to a generic
$g_i^2/(4\pi)$ or $h_f^2/(4\pi)$.

In the Yukawa sector there is no symmetry reason
for can\-cella\-tions and 
$\beta^{(2)}_{h_f} / \beta^{(1)}_{h_f} = 
\calo[(\alpha/4\pi)$ $(2\mu/M_c)]$. An estimate 
suggests that the one and the two-loop
contributions to $\beta_{h_t}$ have the same size for 
$\mu\approx 10 M_c$. In this particular model, however, 
the sign of $\beta_{h_t}$ is always negative, what 
makes $h_t$ a decreasing function of $\mu$ and widens
the range of energies with perturbative values of 
$h_t^2(\mu/M_c)$. Actually, at two-loops all the gauge 
and the Yukawa couplings 
except for $\alpha_1$ decrease with the scale in such 
a way that the 
product $\alpha \mu$ is approximately constant at 
$\mu \gg M_c$. We obtain that at $\mu=10 M_c$
$\beta^{(2)}_{h_t} / \beta^{(1)}_{h_t} \approx 0.3$.
Whether or not higher order corrections could change
substantially the two-loop result is then unclear: 
a factor of 10 from the diagramatics at the 
next order could make $\beta^{(3)}_{h_t} \approx 
\beta^{(1)}_{h_t}$.

In summary, our analysis shows that in models with 
an extra dimension the one-loop running of Yukawa
couplings 
is clearly insufficient, since higher order corrections
can be just as important at scales few times above $M_c$.
Although this type of large corrections vanish in the 
running of the gauge couplings,
to claim unification one needs to make sure that
$h_t$ stays perturbative up to the unification scale.
In the particular models studied here the negative
sign of the one and two-loop $\beta_{h_t}$
suggests that this is the case. But, since the
expansion parameter is $\calo (0.3)$ much below $M_X$,
gauge unification could still be spoiled by higher order
corrections.

\section*{Acknowledgements} 

The author thanks J.~Prades for discussions. 
This work has was supported by CICYT under contract AEN96-1672
and by the Junta de Andaluc\'\i a under contract FQM-101.

\newpage

\setlength{\unitlength}{1cm}
\begin{figure}[htb]
\begin{picture}(10,21)
\epsfxsize=9.cm
\put(3.5,1.5){\epsfbox{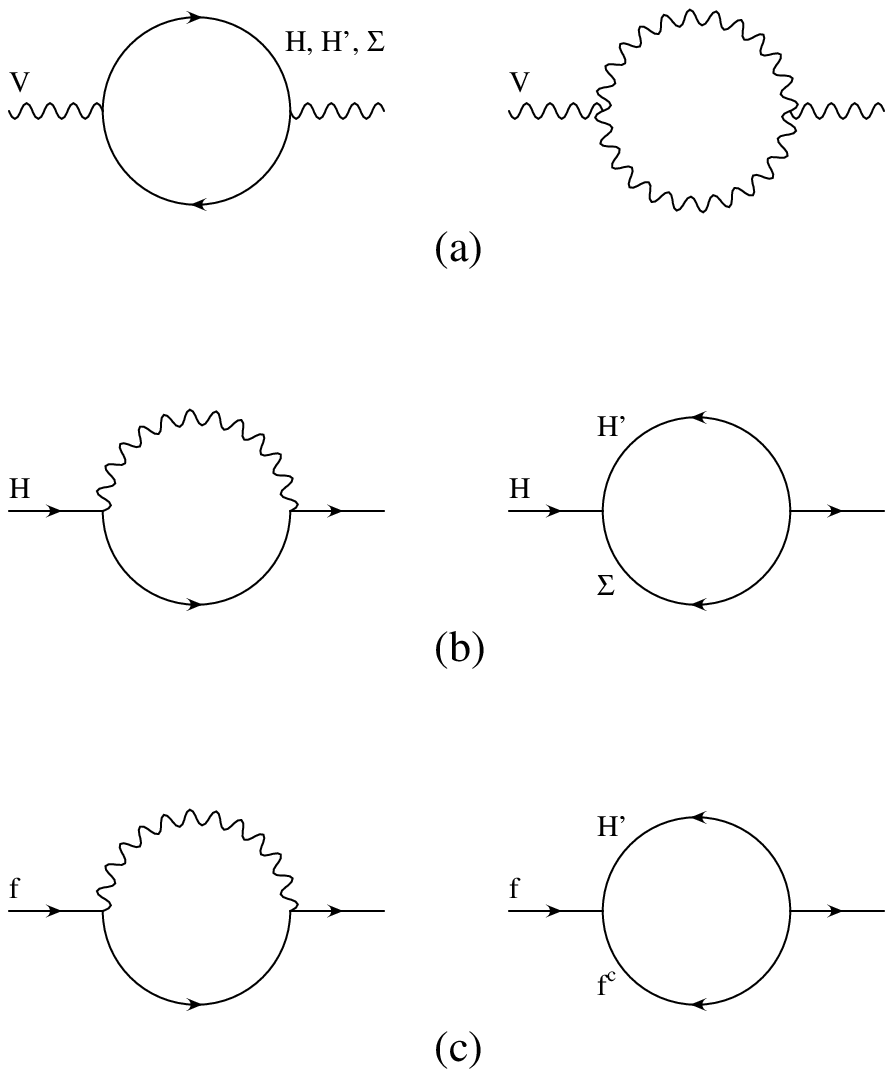}}
\end{picture}
\caption{One-loop contributions to the vector, Higgs and
fermion selfenergies. Solid (wiggle) lines indicate chiral 
(vector) supermultiplets. Ghost contributions are not
included. The two diagrams in Fig.~1(b) cancel.
\label{Fig. l}}
\end{figure}

\newpage

\setlength{\unitlength}{1cm}
\begin{figure}[htb]
\begin{picture}(10,21)
\epsfxsize=9.cm
\put(3.5,0.5){\epsfbox{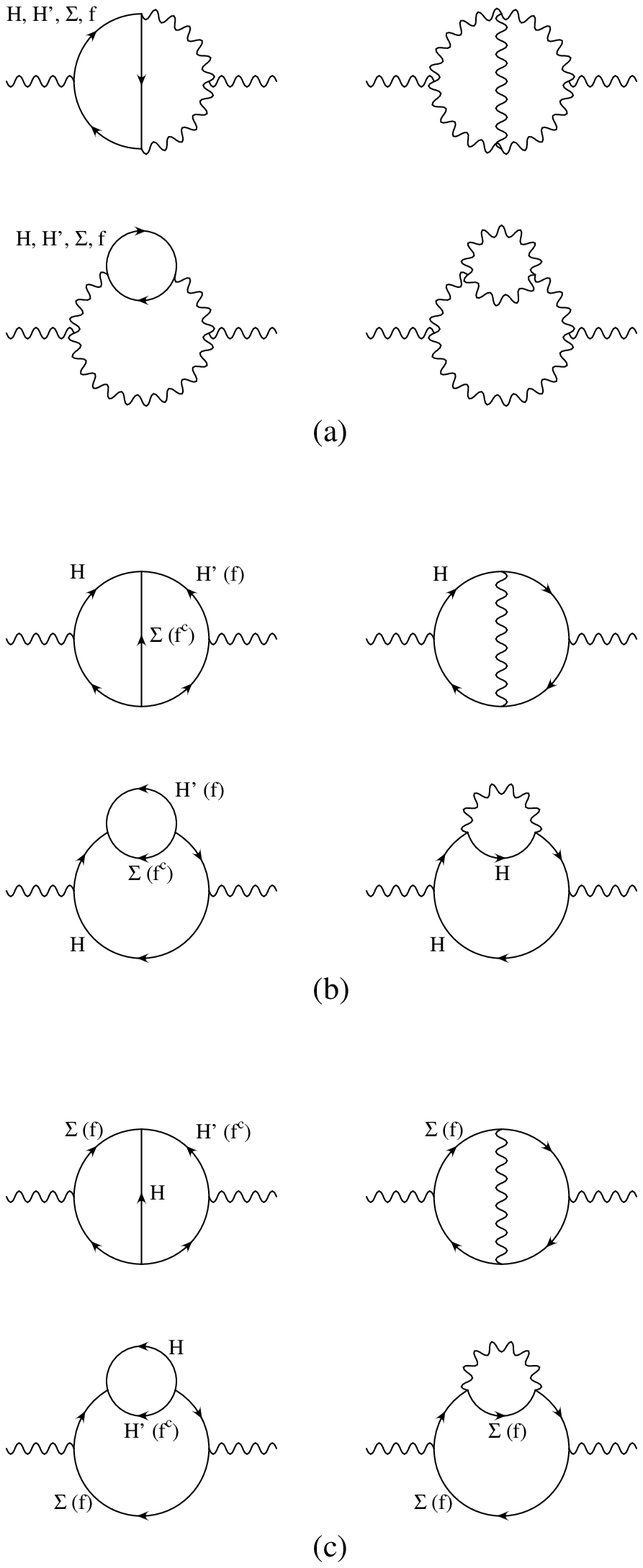}}
\end{picture}
\caption{Two-loop contributions ($\propto \mu/M_c$) to the vector
selfenergy. Counterterm and ghost contributions are not
included.
\label{Fig. 2}}
\end{figure}

\newpage

\setlength{\unitlength}{1cm}
\begin{figure}[htb]
\begin{picture}(10,21)
\epsfxsize=9.cm
\put(3.5,1.5){\epsfbox{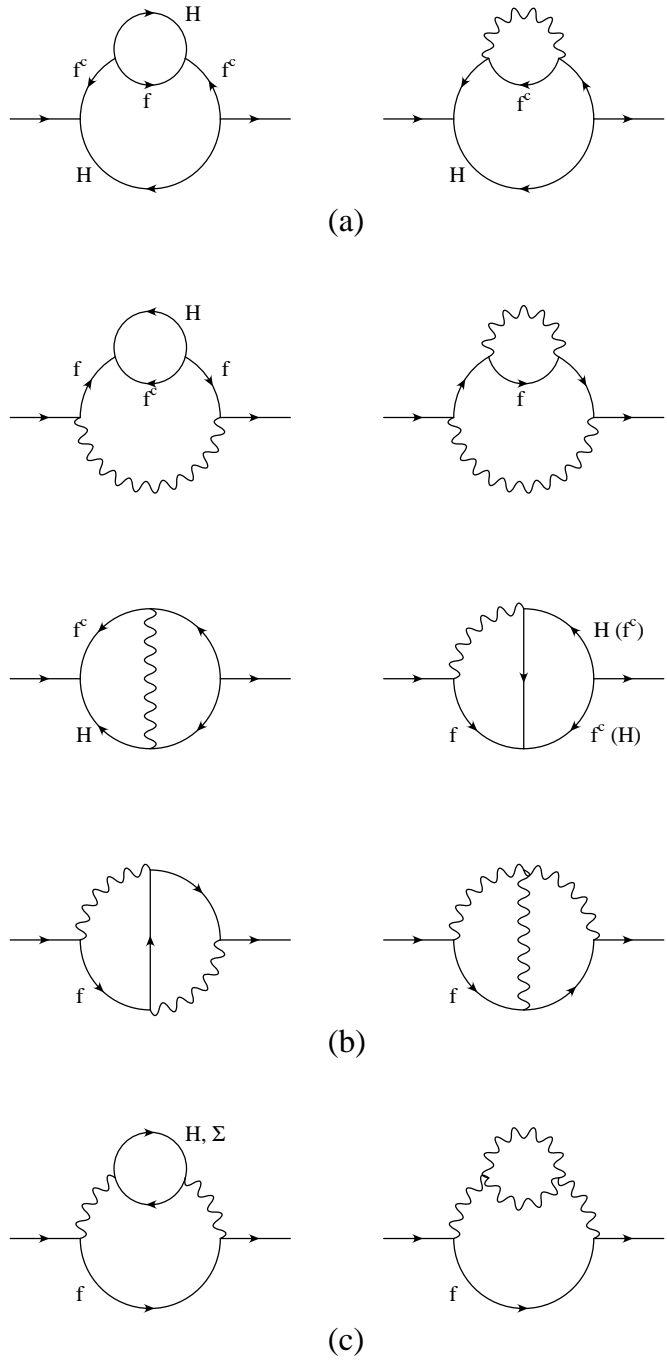}}
\end{picture}
\caption{Two-loop contributions ($\propto \mu^2/M_c^2$) to the 
fermion selfenergies. 
\label{Fig. 3}}
\end{figure}

\newpage

\setlength{\unitlength}{1cm}
\begin{figure}[htb]
\begin{picture}(10,18)
\epsfxsize=25.cm
\put(-4.5,-20){\epsfbox{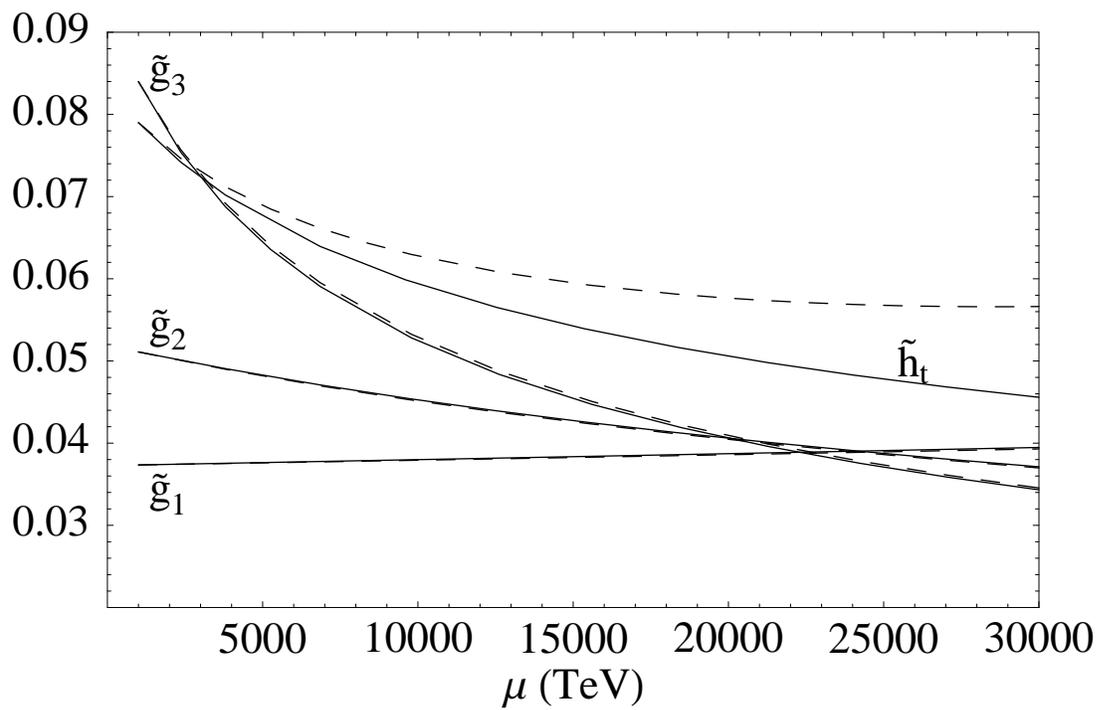}}
\end{picture}
\caption{One-loop (dashed) and two-loop (solid) running
of gauge and Yukawa couplings for the minimal model with 
$\tan\beta=2$ and $M_c=1$ TeV.
\label{Fig. 4}}
\end{figure}

\newpage

\setlength{\unitlength}{1cm}
\begin{figure}[htb]
\begin{picture}(10,18)
\epsfxsize=25.cm
\put(-4.5,-20){\epsfbox{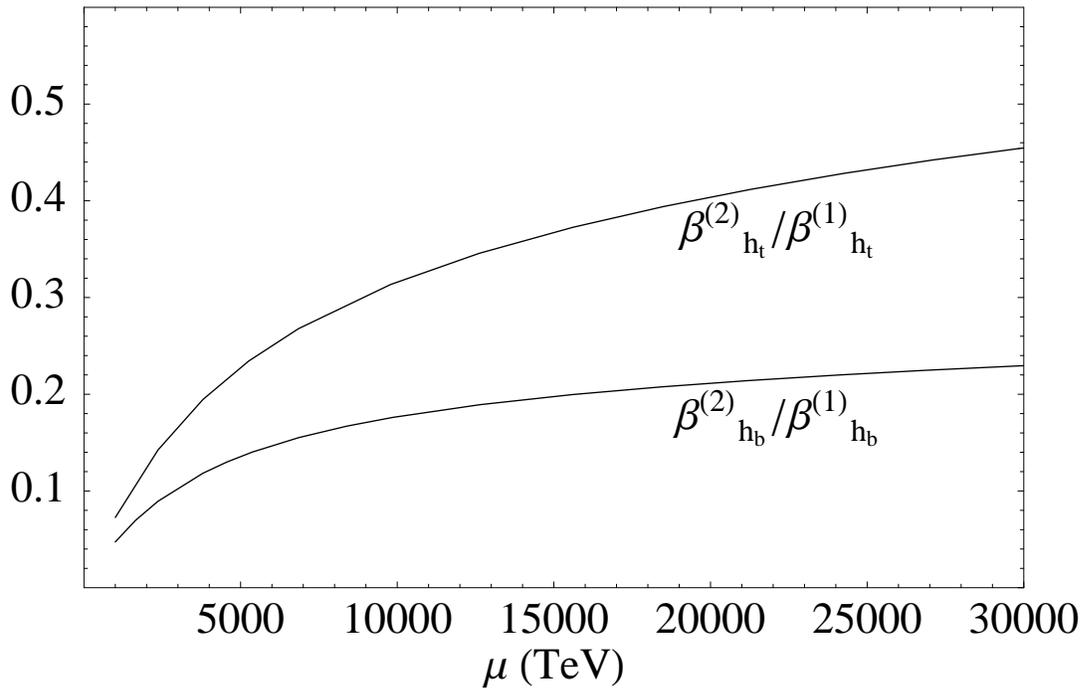}}
\end{picture}
\caption{Absolut value of 
$\beta^{(2)}_{\th_t} / \beta^{(1)}_{\th_t}$ 
and $\beta^{(2)}_{\th_b} / \beta^{(1)}_{\th_b}$ for 
the model in Fig.~4.
\label{Fig. 5}}
\end{figure}

\newpage

\setlength{\unitlength}{1cm}
\begin{figure}[htb]
\begin{picture}(10,18)
\epsfxsize=25.cm
\put(-4.5,-20){\epsfbox{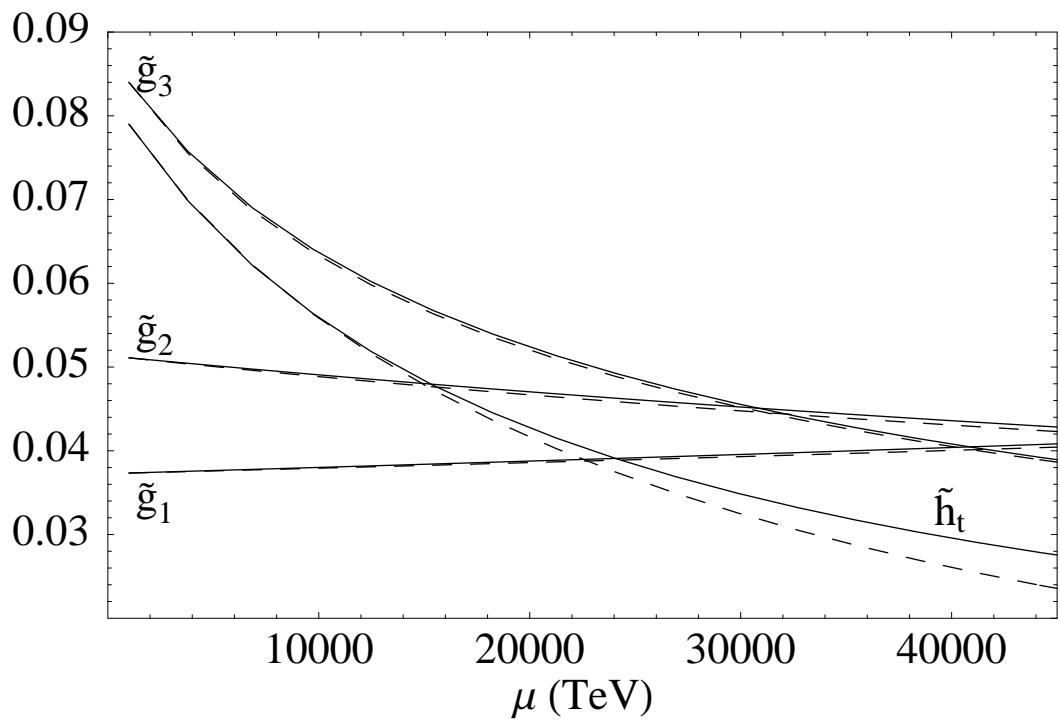}}
\end{picture}
\caption{One-loop (dashed) and two-loop (solid) running
of gauge and Yukawa couplings for the model compactified
on $S^1/Z_2$ with 
$\tan\beta=2$ and $M_c=1$ TeV.
\label{Fig. 6}}
\end{figure}


\newpage   
\begin{thebibliography}{99}   
%
\bibitem{ark98}
N.~Arkani-Hamed, S.~Dimopoulos and G.~Dvali, 
\PLB{429}{98}{263}; 
I.~Antoniadis, N.~Arkani-Hamed, S.~Dimopoulos and G.~Dvali, 
\PLB{436}{98}{257}.
%
\bibitem{kak98a} 
I.~Antoniadis, \PLB{246}{90}{377};
I.~Antoniadis and K.~Benakli, \PLB{326}{94}{69};
Z.~Kakushadze and S.-H.H.~Tye, 
\NPB{548}{99}{180}.
%
\bibitem{die98}
K.R.~Dienes, E.~Dudas, T.~Gherghetta, 
\PLB{436}{98}{55};
\NPB{537}{99}{47}.
%
\bibitem{kub99}
J.~Kubo, H.~Terao, G.~Zoupanos,  hep-ph/9910277.
%
\bibitem{soh85}
M.~Sohnius, \PR{128}{85}{39}.
%
\bibitem{ghi98} 
D.~Ghilencea and G.G.~Ross, \PLB{442}{98}{165};
Z.~Kakushadze, \NPB{548}{99}{205};
S.A.~Abel and S.F.~King, \PRD{59}{99}{095010}; 
T.~Kobayashi, J.~Kubo, M.~Mondrag\'on and G.~Zopaunos,
\NPB{550}{99}99; 
C.D.~Carone, \PLB{454}{99}{70};  
P.H.~Frampton and A.~Ra\v{s}in, \PLB{460}{99}{313};
A.~Delgado and M.~Quir\'os, \NPB{559}{99}235;
D.~Dumitru and S.~Nandi, hep-ph/9906514.
%
\bibitem{che99}
H-C.~Cheng, B.A.~Dobrescu and C.T. Hill, 
hep-ph/9906327.
%
\bibitem{pom98} A.~Pomarol and M.~Quir{\'o}s, \PLB{438}{98}{255};  
A.~Delgado, A.~Pomarol and M.~Quir{\'o}s, hep-ph/9812489.
%
\bibitem{jon84} D.R.T.~Jones and L.~Mezincescu, \PLB{136}{84}{242};
I.~Jack and D.R.T. Jones, \PLB{333}{94}{372}.
%
\bibitem{kak99} 
Z.~Kakushadze and T.R.~Taylor, hep-th/9905137.
%
\bibitem{how83} P.S.~Howe, K.S.~Stelle and P.C.~West, 
\PLB{124}{83}{55}.
%
\bibitem{par84} A.~Parkes and P.~West, \PLB{138}{84}{99}.
%
\bibitem{wes84} P.~West, \PLB{137}{84}{371}.
%


\end{thebibliography}
\end{document}